\documentclass[preprint]{revtex4}
\usepackage{amsfonts}
\usepackage{amsmath}
\usepackage{amssymb}
\usepackage{graphicx}
\usepackage{booktabs}
\providecommand{\U}[1]{\protect\rule{.1in}{.1in}}

\begin{document}

%%%%%%%%%%%%%%%%%%%%%%%%TITLE%%%%%%%%%%%%%%%%%%%%%%%%
\title{
%\vskip-2.5truecm
%\rightline{\small{\tt ULB-TH/08-15}}
%\vskip2.5truecm
{An Attempt to Derive the Expression of the Constant in the Thermodynamic Uncertainty Relations by a Statistical Model for a Quasi-Ideal Nano-Gas}
}
%%%%%%%%%%%%%%%%%%%%%%END_TITLE%%%%%%%%%%%%%%%%%%%%%%%

%%%%%%%%%%%%%%%%%%%%%%ADDRESSES%%%%%%%%%%%%%%%%%%%%%%
\author{ Giorgio SONNINO}
\email{giorgio.sonnino@ulb.be}
\affiliation{
Universit{\'e} Libre de Bruxelles (U.L.B.)\\
Department of Theoretical Physics and Mathematics\\
Campus de la Plaine C.P. 231 - Bvd du Triomphe\\
B-1050 Brussels - Belgium\\
Email: giorgio.sonnino@ulb.be
}

%%%%%%%%%%%%%%%%%%%%END_ADDRESSES%%%%%%%%%%%%%%%%%%%%%

%%%%%%%%%%%%%%%%%%%%%%ABSTRACT%%%%%%%%%%%%%%%%%%%%%%%
\begin{abstract}
In recent work, we have shown that fundamental quantities such as the total entropy production, the thermodynamic variables conjugate to the thermodynamic forces, and the Glansdorff-Prigogine’s dissipative variable may be discretized at the mesoscopic scale. The Canonical Commutation Rules (CCRs) valid at the mesoscopic scale have been postulated and the measurement process consists of determining the eigenvalues of the operators associated with the thermodynamic quantities. The essence of that work was to analyze the consequences of these postulates. This letter aims to understand the physical nature of the new constant $\beta$ introduced in the CCRs. In particular, two questions are still open. Is $\beta$ a new fundamental constant? Does the constant in the CCRs correspond to the lowest limit? We shall tackle the problem starting from the simplest assumption: the expression of $\beta$ can be derived from a simple model (a heuristic model) for nano-gas and studied through the tools of classical statistical physics. We will find that the theoretical value is very close to the experimental one. We shall not go further in our interpretation; we shall limit ourselves simply by noticing that according to our model, the constant $\beta$ does not appear to be a new fundamental constant but corresponds to the minimum value.

\end{abstract}

\maketitle
%%%%%%%%%%%%%%%%%%%%%END_ABSTRACT%%%%%%%%%%%%%%%%%%%%%

%%%%%%%%%%%%%%%%%%%%%%TEXT_PAPER%%%%%%%%%%%%%%%%%%%%%%

One of the main objectives of the Brussels School of Thermodynamics, created by Th{\'e}phile De Donder and Ilya Prigogine, was to investigate systems on a mesoscopic scale to discover the fundamental laws governing them. Inspired by this goal and encouraged by recent experimental results, we established the Canonical Commutation Rules (CCRs) valid at the mesoscopic scale in a previous work. The CCRs show that the closer we get to the mesoscopic level, the more indeterminate the simultaneous measurement of the canonically conjugate variables \cite{sonnino}. In \cite{sonnino}, quantity $\beta$ enters the CCRs. Fourier's theorem provides the lowest limit to the product of the two variances: time and the entropy production rate. This letter investigates the physical origin of constant $\beta$ entering the CCRs. We start with the simplest assumption: \textit{by a statistical model for nano-gases we can derive the expression for $\beta$}. To this end, we adopt a (very) simple heuristic model for nano-gas with the following assumptions
\vskip0.2truecm

\noindent 1) The limit case is reached when the distance between the molecules of the nano-gas is (approximately) twice the Bohr radius ($r_B$).

\noindent 2) The spherical-molecule model is adopted. Beyond the Heisenberg principle, classical statistical physics applies.
\vskip0.2truecm

\noindent Let us consider the first assumption. In Schrödinger's quantum-mechanical theory of the hydrogen atom, the Bohr radius represents the most probable distance between the electron and nucleus of a hydrogen atom at its ground state. The Bohr radius is often used to estimate the size scale where quantum effects become significant. In most scenarios where molecular collisions are involved, if the velocities of the molecules are considered sufficiently high, the model using $4 \pi r_B^2$ as the effective interaction area in classical scattering theory is valid as a first approximation. This model is valid at a first approximation for describing classical effects at a mesoscopic scale for the following reasons:

\noindent i) A cross-section greater than $4 \pi r_B^2$ signifies a relatively large interaction area between the colliding molecules. In this scenario, the collision can be visualized as a more direct, head-on encounter, resembling the interaction of classical hard spheres.

\noindent ii) When the molecules have high velocities, their de Broglie wavelength (wavelength associated with their wave nature) becomes smaller. Consequently, the quantum mechanical effects like wave-particle duality become less prominent. This allows classical mechanics to provide a more reasonable approximation.

\noindent iii) Mesoscopic scale refers to a length scale larger than atoms/molecules but smaller than the bulk material. This avoids complications at very small (quantum regime) or large (macroscopic material properties) scales.

\noindent However, as with any approximation, it is important to recognize its limitations and consider the specific context in which it is being applied. The effectiveness of the $4 \pi r_B^2$ cutoff depends on the specific molecules. It might not be accurate for complex molecules with intricate shapes. Furthermore, there is no sharp dividing line between classical and quantum behavior. Even with a large cross-section, quantum effects might still be relevant for very high precision or extremely low velocities. Finally, with the limitations mentioned above, we can say that, if the velocity of the molecules is sufficiently high, the model using $4\pi r_B^2$ as the effective interaction area in classical scattering theory is valid as a first approximation. This approximation provides a simple and convenient way to estimate the cross-section of molecular collisions and can be useful in many practical situations, especially when quantum effects are not dominant and relativistic effects can be ignored \cite{atkins}, \cite{wyatt}, \cite{lishchuk}.

\begin{figure}
\includegraphics[width=7cm]{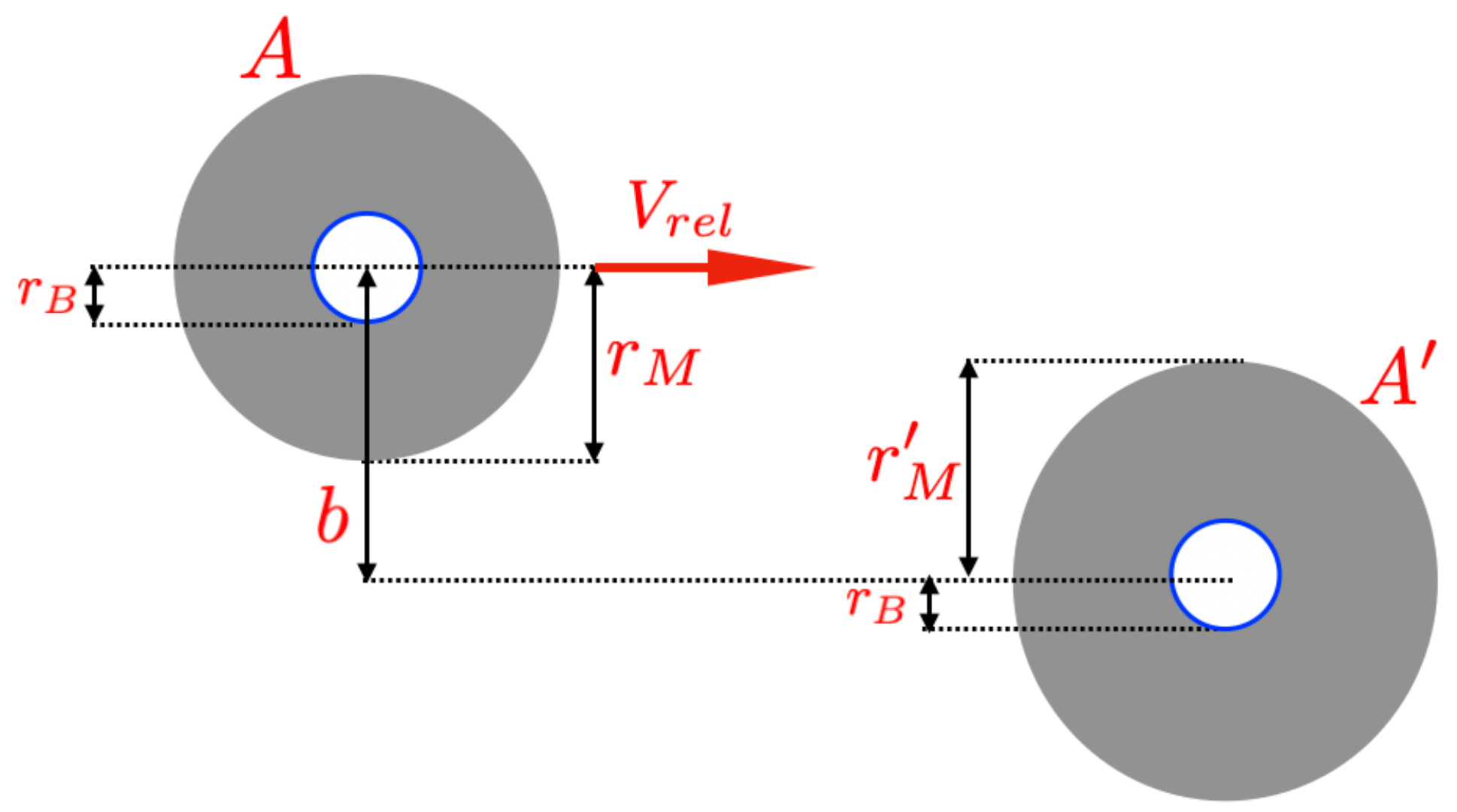}
\caption {{\bf Collision between two spherical molecules of a quasi-ideal nano-gas}. \textit{A collision between two spherical molecules of radius $r'_M$ and $r_M$ with relative velocity $V_{rel}$. The \textit{impact parameter} $b$ is the distance between the two centers of the molecules. In our heuristic model, the molecules are identical, so $r'_M=r_M$ and the volume occupied by a molecule is $V_M= 4/3\pi r_M^3$. The model assumes that if $b\leq2r_B$ the two molecules collide, they "feel" the Heisenberg principle. The limit case is reached when the distance between the two molecules is $b=2r_B$ (so, the cross-section is $\sigma_{cs}=\pi b^2=4\pi r_B^2$) and the perfect packing is reached (i.e., $n=V^{-1}_M$}).}
\label{modello}
\end{figure}

\noindent Another issue comes from nanomaterials not being limited to just one kind of molecule. They can be built by organic molecules (polymers like ADN or synthetic plastics, etc.), inorganic molecules (e.g., metals like gold or silicon nanoparticles, etc.), or compounds (e.g., many nanomaterials are metal oxides, like titanium dioxide, or complex compounds with unique structures). Furthermore, in some cases, the nanomaterial might not be individual molecules but clusters of atoms or molecules held together by specific forces. However, this work only aims to provide a starting point for understanding the minimum product $\Delta t\Delta\sigma$, with $\Delta\sigma$ denoting the uncertainty related to the measurements of the entropy production strength. Fig.~\ref{modello} shows the classical spherical model to represent molecules in nano-gas. In our case, the impact parameter $b$ is half of Bohr's radius and all the molecules of the nano-gas are equal (so, $r'_M=r_M$ with $r_M$ denoting the molecular radius).
%%%%%%%%%%%%%%%%%%%%%%%%%%%%%%%%%%%%%%%%%%%%%%%%%%%%%%%%%%%%%%%%%%%%%%%%%%%%%%%
%%%%%%%%%%%%%%%%%%%%%%%%%%%%%%%%%%%%%%%%%%%%%%%%%%%%%%%%%%%%%%%%%%%%%%%%%%%%%%%
\noindent At thermal equilibrium (i.e., the temperature of the system constant), the (classical) uncertainty related to the measurement of the entropy production strength $\sigma$ reads \footnote{Indeed, from the first thermodynamic law, we have $dE=TdS$ (no work on or by the system), with $dE$ and $dS$ denoting the variation of the energy and the variation of the total entropy of the system, respectively. From Prigogine's law, $dS$ is given by two contributions: a piece due to reversible transformations and a piece due to dissipative processes \cite{prigogine}. In our case, dissipation arises only from collisions and there is no reversible entropy flow during the collision process. So, $dS=d_IS$ with $d_IS$ denoting the variation of the entropy production of the system. The entropy production strength $\sigma$ is given by $\sigma=d_IS/dt$. In scattering processes, if the temperature of the system remains constant and assuming that the smallest time steps are equal to the collision time $\tau$, the variation of the entropy production strength reads $\Delta\sigma\sim\Delta E/(T\tau)$.}:
\begin{equation}\label{m2}
\Delta\sigma\sim\frac{\Delta E}{T\tau}
\end{equation}
\noindent with $\tau$ denoting the collision time, $T$ the temperature of the system, and $\Delta E$ the energy uncertainty of the classical system, respectively. Collisions among molecules within a system induce a dissipative process. Indeed, kinetic energy can be transferred between molecules in a collision, and some of this energy can be converted into other forms such as heat due to friction or deformation. This conversion of kinetic energy into other forms typically leads to a dissipation of energy within the system. In a gas, when molecules collide, they may exchange kinetic energy, and some of this energy may be converted into rotational or vibrational energy, increasing the overall internal energy of the system. This process is often associated with an increase in entropy production, which is a hallmark of dissipative processes in thermodynamics. So, while collisions themselves are essential for maintaining the dynamics and equilibrium of a system, they contribute to the dissipation of energy within the system. In classical statistical mechanics, the equipartition theorem relates the temperature of a system to its average energies. In particular, it predicts that the gas of molecules in thermal equilibrium at temperature $T$ has an average translational kinetic energy of $3/2 k_B T$ where $k_B$ is the Boltzmann constant:
\begin{equation}\label{m3}
\frac{1}{2}m_M{\bar{v^2}}=\frac{3}{2}k_BT
\end{equation}
\noindent with $\bar {v^2}$ and $m_M$ denoting the mean square speed and the molecule mass, respectively. The statistical theory of gases provides us with the expression of the collision time \cite{reif}:
\begin{equation}\label{m4}
\tau=\frac{1}{n\sigma_{cs}{\bar V_{rel}}}
\end{equation}
\noindent where $\sigma_{cs}=\pi b^2=4\pi r_B^2$ is the cross section. $n$ and ${\bar V_{rel}}$ are the number of molecules for unit volume and the mean relative velocity, respectively. The choice of $\sigma_{cs}=4\pi r_B^2$ is reasonable for representing the effective interaction area of a molecule. This approximation is often used in classical scattering theory \cite{goldstein}. ${\bar V_{rel}}$ is linked to the root mean square speed by the relation $\overline{V^2_{rel}}=2{\bar{v^2}}$ \cite{reif}. If we do not make a distinction between the mean of the square and the square of the mean, we have $\bar{V}_{rel}\simeq \sqrt{2{\bar{v^2}}}$ \cite{reif}. By combining all these expressions we get 
\begin{equation}\label{m5}
\Delta t\Delta\sigma= (\Delta t\Delta E)\frac{3 n\sigma_{cs}}{\sqrt{2}m_M\sqrt{{\bar{v^2}}}}k_B\geq \frac{\hbar}{2} \frac{3 n\sigma_{cs}}{\sqrt{2}m_M\sqrt{{\bar{v^2}}}|_{Max.}}k_B
\end{equation}
\noindent In Eq.~(\ref{m3}) we have taken into account the Heisenberg uncertainty principle:
\begin{equation}\label{m1}
\Delta t\Delta E\geq\frac{\hbar}{2}
\end{equation}
\noindent with $\hbar$ denoting the reduced Planck constant. Here, $\Delta t$ and $\Delta E$ are the uncertainties related to the time and energy measurements on a system, respectively. Let us now consider the limit case. According to postulate 2), the limit case is reached when the distance between the centers of the molecules $b$ is twice the Bohr radius. Below this limit, scattering processes are governed by quantum laws and no longer by the classical ones. Scattering processes produce entropy and lead to various forms of energy exchange and redistribution within the material, resulting in changes in the system's overall entropy. Additionally, nano-gas often operate under non-equilibrium conditions, where external forces (like electric fields, temperature gradients, or mechanical stress) drive transport processes. These non-equilibrium conditions lead to entropy production as the system strives to return to equilibrium. According to our (raw) model, molecules collide with each other with a minimal impact factor when
\begin{equation}\label{m6}
n\sigma_{cs}=3\eta\frac{r_B^2}{r_M^3}
\end{equation}
\noindent with $\eta$ denoting the \textit{packing fraction} and $r_M$ the radius of the average molecule of the nanomaterial, respectively. A perfect packing corresponds to $\eta=1$. The product $\Delta t\Delta\sigma$ is smaller the smaller the average distance between molecules. Indeed, minimizing the average distance between molecules, as achieved in the limit case of perfect packing ($\eta=1$), reduces the uncertainty in entropy production strength ($\Delta\sigma$): densely packed molecules lead to a more predictable and less uncertain behavior in terms of entropy production. Hence, achieving perfect packing ($\eta=1$) minimizes the average distance between molecules and reduces the uncertainty in entropy production strength, leading then to a lower product $\Delta t\Delta\sigma$. Finally, the limit case corresponds to a perfect packing ($\eta =1$) (all available space occupied by molecules) with $\sigma_{cs}=\pi b^2=4\pi r_B^2$. We are now in a position to get a rough estimation of $\beta$. Indeed,
\begin{equation}\label{m7}
\Delta t\Delta\sigma\geq\frac{9}{2\sqrt{2}}\frac{\hbar}{m_ecr_B}\left(\frac{r_B}{r_M}\right)^3\left(\frac{m_e}{m_M}\right)k_B=\frac{9}{2\sqrt{2}}\alpha\left(\frac{r_B}{r_M}\right)^3\left(\frac{m_e}{m_M}\right)k_B=\beta k_B
\end{equation}
\noindent with $m_e$ denoting the electron mass, $c$ the speed of light, and $\alpha$ the fine-structure constant, respectively. Note that although the mass of the electron appears in Eq.~(\ref{m7}) it plays no role since the product $\alpha m_e$ is independent of the mass of the electron. However, as we will see shortly, it is important from the physical point of view to put this quantity in evidence. From Eq.~(\ref{m7}) we get,
\begin{equation}\label{m8}
\beta=\frac{9}{2\sqrt{2}}\alpha\left(\frac{V_B}{V_M}\right)\left(\frac{m_e}{m_M}\right)=\frac{9}{2\sqrt{2}}\alpha\chi
\end{equation}
\noindent where $V_B/V_M$ represents the relative size of the Bohr volume (volume of an atom based on the Bohr radius) to the volume occupied by a single molecule and the ratio $m_e/m_M$ compares the mass of an electron to the mass of a molecule composing the material, respectively. Parameter $\chi\equiv (V_Bm_e)/(V_Mm_M)$ denotes the \textit{Bohr-Molecular Ratio} (BMR) \footnote{This name comes from the fact that it involves the \textit{Bohr sphere mass} (BSM) and the molecular properties. $\chi$ compares the atomic scale (the product of the Bohr sphere and the mass of the electron) with the molecular scale (the \textit{molecule's bulk properties}, volume and mass).}.

\noindent {\bf Discussion}

\noindent The Bohr radius is about $r_B=5.29\times 10^{-11}$ meters. The diameter of molecules in nanomaterial can vary significantly. Our calculation assumed a perfect sphere for the molecule of nano-gas (which is not always the case). So, while the actual ratio might depend on the specific nano-gas, a molecular radius of the order $r_M\sim 5\times 10^{-10}$ meters and $r_b/r_M \sim 0.1$ is a reasonable estimate for the order of magnitude \cite{reif}. So, $V_B/V_M$ is of the order of $10^{-3}$. As discussed before, the mass of molecules in nanomaterials can vary significantly. However, the mass is likely several thousand to millions of times heavier than an electron for many common molecules. Therefore, saying the ratio between the electron mass and an average molecule's mass is about the electron-to-proton mass ratio captures the significant difference in scale between these two quantities. To determine the ratio between the mass of an electron and an average molecule in a system on a mesoscopic level, we need to consider the mass of the entire molecule, which includes the masses of all the atoms it contains, along with their constituent protons, neutrons, and electrons. The mass of an electron is approximately $m_e=9.109\times10^{-31}$ kilograms. A nanomaterial can be composed of a wide range of elements and have diverse molecular structures, so it is challenging to give a single specific ratio without knowing the exact composition of the nanomaterial. One example is graphene, a two-dimensional nanomaterial composed of a single layer of carbon atoms arranged in a hexagonal lattice. In graphene, the ratio between the effective mass of electrons (which behaves differently than the free electron mass in a vacuum due to interactions with the crystal lattice) and the molecular mass (carbon's atomic mass) is approximately 1/3000. This property of graphene contributes to its unique electronic properties, such as high electron mobility, which is important for various applications in electronics, sensors, and other fields. Experimental techniques such as scanning tunneling microscopy (STM), atomic force microscopy (AFM), Raman spectroscopy, and others can be used to study the properties and dynamics of graphene-based materials at the nanoscale, including the effects of scattering processes on entropy production. For our purposes, it is reasonable to assume that $m_e/m_M\sim 10^{-3}-10^{-4}$. Finally, the Bohr-Molecular Ratio $\chi=(V_B m_e)/(V_Mm_M)$ is of the order of $10^{-6}-10^{-7}$. This suggests that the quantum effects described by the Bohr radius are not dominant on the mesoscopic scale and that classical physics can provide an adequate description of nano-gas behavior. Indeed,

\noindent i) $V_B/V_M$ is proportional to the cube of the Bohr radius. At larger length scales, such as the mesoscopic level, the volume of interest ($V_M$) is significantly larger than the Bohr volume ($V_B$).

\noindent ii) If the ratio $m_e/m_M$ is much smaller than unity, it suggests that the mass of the electron is negligible compared to the mass of the molecules. At the mesoscopic level, where we deal with large numbers of atoms and molecules, the mass of the molecules dominates, making the ratio $m_e/m_M$ very small.

\noindent iii) The BMR $\chi= (V_Bm_e)/(V_Mm_M)$ compares the scale of fundamental atomic properties (Bohr radius and electron mass) to the scale of molecular properties (molecular volume and mass). The smallness of $\chi$ indicates the disparity in scales between the atomic level and the molecular level. So, the result that the discretization constant $\beta$ is proportional to the ratio $\chi$ makes sense. This formulation effectively captures the disparity between atomic and molecular scales, providing a meaningful quantization of entropy production at the mesoscopic level. The use of fundamental constants and molecular properties supports the robustness of this result within the framework of mesoscopic thermodynamics.

\noindent By plugging the values $\alpha=1/137$, $V_B/V_M\sim 0.3$, and $m_e/m_M=1/3000$ into Eq.~(\ref{m8}) we get $\beta_{Theor.}\sim 10^{-8}$ which aligns with the experimental value found in \cite{roldan}: $\beta_{Exp.}\sim 1.2\times 10^{-8}$.

\section{Concluding Remarks}
In \cite{sonnino} we have shown that fundamental quantities such as the total entropy production, the thermodynamic variables conjugate to the thermodynamic forces, and the Glansdorff-Prigogine’s dissipative variable may be discretized at the mesoscopic scale. The aim of \cite{sonnino} was not to present a model for calculating $\beta$, but to explore the consequences of the fact that the commutator between time and entropy production strength is different from zero. This letter addresses two crucial questions:

\noindent 1) \textit {Is $\beta$ a universal constant?}

\noindent 2) \textit {Does $\beta$ correspond to the lowest limit?}

\noindent To answer these questions we proceed step by step. This first investigation started by assuming that it is possible to deduce the expression of $\beta$ through a very simple model for nano-gas that satisfies the laws of classical statistical physics. Other kinds of investigations will follow in future work. According to our model, the answer to the first question is "no" and is affirmative for the second one. Furthermore, the value provided by the theoretical expression is very close to the experimental one found in \cite{roldan}. However, while the reasoning presented addresses some relevant aspects, there are several limitations to consider. For instance, the assumption of perfect packing ($\eta=1$) may not always be feasible in real-world scenarios. Achieving perfect packing in nanomaterials may be challenging due to molecular size variations, steric hindrance, and thermal fluctuations. Real nanomaterials often exhibit imperfect packing, affecting their properties and behavior. Our reasoning focused on minimizing the average distance between molecules to reduce the uncertainty in entropy production strength. However, it simplifies the role of interactions between molecules. In reality, the nature and strength of molecular interactions play a crucial role in determining the system's entropy production and overall behavior. Thus, we conclude that Eq.~(\ref{m8}) provides only a rough estimate of $\beta$. Eq.~(\ref{m8}) confirms that it makes sense to postulate that the entropy production strength and the thermodynamic variables conjugated to the thermodynamic forces can be discretized on a mesoscopic scale. In \cite{barato} the authors analyzed biomolecular systems like molecular motors or pumps, transcription and translation machinery, and other enzymatic reactions. They showed that in a steady state the dispersion of observables, is subject to an uncertainty relation. More specifically, by investigating the simplest case of a nonequilibrium chemical reaction catalyzed by an enzyme, where a single step is interpreted as the completion of an enzymatic cycle, they showed that the following inequality constrains the product (time-entropy production strength):
\begin{equation}\label{m8}
t \sigma\geq 2\frac{(k^+-k^-)^2\tau_{step}}{k^++k^-}k_B=\beta k_B
\end{equation}
\noindent where the steps to the right happen with a rate $k^+$, those to the left with a rate $k^-$, respectively, and $\tau_{step}$ is the elementary time step. The elementary time step in the context of a biased random walk for enzymatic reactions is typically considered to be the mean time it takes for the enzyme to complete one full cycle of reaction, encompassing all the steps mentioned above. Thus, we can state that the operators associated with the canonically conjugate variables do not commute with each other and their commutators never vanish.

%%%%%%%%%%%%%%%%%%%%%END_TEXT_PAPER%%%%%%%%%%%%%%%%%%%%

%%%%%%%%%%%%%%%%%%%%%BIBLIOGRAPHY%%%%%%%%%%%%%%%%%%%%%%

%%%%%%%%%%%%%%%%%%%%%END_BIBLIOGRAPHY%%%%%%%%%%%%%%%%%%%


\begin{thebibliography}{alpha}

\bibitem{atkins} P. W. Atkins and R. S. Friedman, 2005, \textit{Molecular Quantum Mechanics}, Oxford University Press Inc., New York.

\bibitem{wyatt} R. E. Wyatt and K. Michielsen, International Journal of Modern Physics B, {\bf 17}(29) (2003).

\bibitem{lishchuk} S. V. Lishchuk, V. V. Kozlov, and A. D. Kudryavtsev, International Journal of Quantum Chemistry, {\bf 102}(4) (2005).

\bibitem{prigogine} I. Prigogine, 1954, \textit{Thermodynamics of Irreversible processes}, (John Wiley \& Sons). 

\bibitem{sonnino} G. Sonnino, \textit{Uncertainty Relations in Thermodynamics of Irreversible Processes on a Mesoscopic Scale}, submitted to publication to Physica E (2024).

\bibitem{goldstein} H. Goldstein, C. P. Poole Jr., and J. L. Safko, 2011, \textit{Classical Mechanics}, Published by Dorling Kindersley (India) Pvt. Ltd, licensees of Pearson Education in South Asia.

\bibitem{reif} F. Reif, Fundamentals of Statistical and Thermal Physics, Waveland Press, Inc., Long Grove Illinois, 1964 (reissue 2009).

\bibitem{roldan} E. Rold{\'a}n, \textit{et al.}, New J. Phys., {\bf 23}, 083013 (2021).

\bibitem{barato} A. C. Barato and U. Seifert, Phys. Rev. Lett., {\bf 114}, 158101 (2015).

\end{thebibliography}
\end{document}